\documentclass[prb,twocolumn,showpacs]{revtex4}
\usepackage{graphicx}
\begin{document}
\title{Surface magnetic canting in a ferromagnet}
\author{P. Betti$^{1}$}
\author{M. G. Pini$^{2}$}
\email{mgpini@ifac.cnr.it}
\author{A. Rettori$^{1}$}
\email{rettori@fi.infn.it} 
%\homepage[]{Your web page}
%\thanks{}
%\altaffiliation{}
\affiliation{$^1$Dipartimento di Fisica, Universit\`a di Firenze and 
Istituto Nazionale per la Fisica della Materia,
Unit\`a di Firenze, Via G. Sansone 1, I-50019 Sesto Fiorentino, Italy\\
$^2$Istituto di Fisica Applicata ``Nello Carrara",
Consiglio Nazionale delle Ricerche, Via Panciatichi 56/30,
I-50127 Firenze, Italy
}
\date{\today}
\begin{abstract}
The surface magnetic canting (SMC) of a semi-infinite film
with ferromagnetic exchange interaction and competing bulk 
and surface anisotropies is investigated via a nonlinear 
mapping formulation of mean-field theory previously 
developed by our group [L. Trallori {\it et al.}, 
Int. J. Mod. Phys. B {\bf 10}, 1935-1988 (1996)], and extended 
to the case where an external magnetic field is applied 
to the system. 
When the field $H_{\Vert}$ is parallel to the film plane, the
condition for SMC is found to be the same as that recently
reported by Popov and Pappas [Phys. Rev. B {\bf 64},
184401 (2001)]. The case of a field $H_{\perp}$ applied 
perpendicularly to the film plane is also investigated.
In both cases, the zero-temperature equilibrium configuration is 
easily determined by our theoretical approach.
\end{abstract}
%75.70.Rf Surface magnetism
%75.70.-i Magnetic properties of thin films, surfaces, and interfaces
%75.30.Kz Magnetic phase boundaries
%45.      Classical mechanics of discrete systems
%05.45.-a Nonlinear dynamics and nonlinear dynamical systems
\maketitle

Recently Popov and Pappas\cite{PP} - motivated by their previous valuable 
experimental work\cite{exptPP} on the magnetic properties of an ultrathin 
(1.5 monolayers-thick) Fe film grown on the surface of a 15nm-thick Gd(0001) 
film - determined the zero-temperature phase diagram of a semi-infinite
Heisenberg ferromagnet, with exchange constant $J>0$, 
subject to a surface anisotropy, $K_S$, competitive with 
the bulk one, $K_B$. In the mean-field approximation, since the system
is inhomogeneous only along one direction ({\it i.e.}, the normal 
to the surface), the problem is reduced to consider the energy of 
the one-dimensional model\cite{PP}
\begin{eqnarray}
\label{model}
E&=&-J\sum_{n=1}^{\infty} \cos(\theta_n-\theta_{n+1})
+ K_S~\sin^2 \theta_1 
\cr
&+&K_B \sum_{n=2}^{\infty} \sin^2 \theta_n
-g\mu_B H_{\Vert} \sum_{n=1}^{\infty} \cos\theta_n
%-g\mu_B H_{\perp} \sum_{n=1}^{\infty} \sin\theta_n
\end{eqnarray}
\noindent where $\theta_n$ is the angle formed by the 
classical vector moment of the $n$-th layer
with the film plane. 
To fix ideas, in the following we will assume
$K_B>0$ ({\it i.e.}, the bulk anisotropy 
favours the alignment of the vector 
moments along the film plane, $\theta=0$) and $K_S<0$ 
({\it i.e.}, the surface anisotropy 
favours the alignment along the film normal, $\theta=\pi/2$).
$H_{\Vert}$ denotes an external magnetic field 
applied parallel to the film plane.

Expanding the energy in Eq. (\ref{model}) 
to second order for each $\theta_n$,
$E\approx E_0+{\bf \theta}^{T} A~ {\bf \theta}$, and performing 
a stability analysis, Popov and Pappas\cite{PP} showed that it 
is possible to have surface magnetic canting (SMC) provided that 
\begin{equation}
k_S+1+h_{\Vert}<{1\over {k_B-(k_S-1)}}
\label{PPC}
\end{equation}
where $k_S=2K_S/J$, $k_B=2K_B/J$ and $h_{\Vert}=g\mu_B H_{\Vert}/J$.
Hence it follows that for $k_S<-1-h_{\Vert}$, 
the surface is always canted whatever the value of the
in-plane bulk anisotropy; otherwise
there is SMC for $|K_S|$ exceeding a threshold value which 
depends on $K_B$, $J$ and $H_{\Vert}$.

Surely, the theoretical analysis of Popov and Pappas\cite{PP} is 
quite original and valuable. However, it is worth noting that,
in the special case $H_{\Vert}=0$, a condition for surface 
magnetic canting analogous to Eq.~(\ref{PPC})
was obtained by our group some years ago\cite{nota1} 
using a rather different method: {\it i.e.}, 
a nonlinear mapping formulation of mean-field 
theory.\cite{PRL,JPL,JMMM,PLA,Review,Matteo}
Within this framework, the properties of a 
magnetic film are formulated as a problem in nonlinear dynamics, 
in terms of an area-preserving map, where the surfaces
are introduced as appropriate boundary conditions.
In this Brief Report, using the method developed in Refs. 
\onlinecite{PRL,JPL,JMMM,PLA,Review,Matteo}, not only we obtain 
the condition for surface magnetic canting, but also we
calculate the zero-temperature magnetization profile of the film,
both in the case of the model, Eq.~(\ref{model}), considered by Popov and 
Pappas\cite{PP}, and in the case that a magnetic field is applied 
perpendicularly to the film plane, see Eq. (\ref{modelperp})
below.\cite{nota2}

Let us summarize our method. 
The equilibrium configurations of the semi-infinite
($n \ge 1$) film are obtained from Eq.~(1) by $\theta_n$-derivation
\begin{eqnarray}
\label{mappa}
{{\partial E}\over {\partial \theta_n}}&=&
J\sin(\theta_n-\theta_{n+1})  
-J\sin(\theta_{n-1}-\theta_n)(1-\delta_{n,1})
\cr
&+&K_S \sin(2\theta_n) \delta_{n,1}
+K_B \sin(2 \theta_n) (1-\delta_{n,1})
\cr
&+&g\mu_B H_{\Vert} \sin \theta_n=0
\end{eqnarray}
Introducing the variables $s_n=\sin(\theta_n-\theta_{n-1})$,
in the bulk  ($n \ge 2$) we obtain the nonlinear mapping 
\begin{eqnarray}
\label{mappabulk}
J s_{n+1}&=&J s_n+K_B \sin(2\theta_n)
+g\mu_B H_{\Vert} \sin\theta_n
\cr
\theta_{n+1}&=&\theta_n+\sin^{-1}(s_{n+1})
\end{eqnarray}
while on the surface plane $n=1$ we have
\begin{eqnarray}
J s_2&=&K_S \sin(2\theta_1)+g\mu_B H_{\Vert} \sin\theta_1
\cr
\theta_2&=&\theta_1+\sin^{-1}(s_2)
\end{eqnarray}
The map, Eq.~(\ref{mappabulk}), is area-preserving because 
its Jacobian determinant is 1
\begin{eqnarray}
\label{Jacobian}
{\rm det} {\hat J} &=& {\rm det} \left\lbrack
\begin{array}{l}
	{{\partial s_{n+1}}\over {\partial s_n}}
\; 
	{{\partial s_{n+1}}\over {\partial \theta_n}}
\; 
	\cr
	{{\partial \theta_{n+1}}\over {\partial s_n}}
\; 
	{{\partial \theta_{n+1}}\over {\partial \theta_n}}
\;  
\end{array}\right\rbrack
\cr
&=&{\rm det}
\left\lbrack
\begin{array}{l}
~~~~1~~~~~~~~~
\; 
{\cal S}(\theta_n)
%{{2K_B}\over J}\cos(2 \theta_n)+
%{{g\mu_B H_{\Vert}}\over J}\cos\theta_n
\; 
	\cr
{1\over {\sqrt{1-s_{n+1}^2}}}~~~
\; 
1+
{1\over {\sqrt{1-s_{n+1}^2}}} 
{\cal S}(\theta_n)
%\left\lbrack
%{{2K_B}\over J} 
%\cos(2\theta_n)
%+{{g\mu_B H_{\Vert}}\over J}\cos\theta_n
%\right\rbrack
\;  
\end{array}\right\rbrack 
\end{eqnarray}
where ${\cal S}(\theta_n)=
{{2K_B}\over J}\cos(2 \theta_n)+
{{g\mu_B H_{\Vert}}\over J}\cos\theta_n$.
The trajectories in ($\theta,s$) space are associated with 
equilibrium configurations, while the fixed points of the map 
correspond to uniform ground states of the 
bulk system. For $K_S<0$ and $K_B>0$, the hyperbolic 
fixed point: ($\theta_{\infty},s_{\infty}$)$=$($0,0$) represents 
the energetically stable bulk configuration, while the
elliptic fixed point: ($\theta_{\infty},s_{\infty}$)$=$($\pi/2,0$)
represents an unstable bulk configuration.

Now we observe that, by introducing a fictitious plane $n=0$,
characterized by the angle $\theta_0$ and the variable 
$s_1=\sin(\theta_1-\theta_0)$,
the nonlinear mapping Eq.~(\ref{mappabulk}), valid in the bulk 
($n \ge 2$), can be assumed to hold even for $n=1$,
\begin{eqnarray}
\label{mappageneral}
J s_2&=& J s_1+K_B\sin(2\theta_1)+g\mu_B H_{\Vert} \sin\theta_1
\cr
\theta_2&=&\theta_1+\sin^{-1}(s_2)
\end{eqnarray}
provided that the following boundary condition is satisfied
\begin{equation}
\label{surface}
s_1={{K_S-K_B}\over J} \sin(2\theta_1).
\end{equation}
Thus, among all trajectories in $(\theta,s)$ space obtained from 
Eqs.~(\ref{mappabulk})(\ref{mappageneral}), 
only those which satisfy Eq.~(\ref{surface})
represent equilibrium configurations for the semi-infinite
ferromagnet. 

By linearizing the map in the neighborhood of a
fixed point,\cite{nota3} one is led to solve the eigenvalue equation
\begin{equation}
\lambda^2-{\rm Tr} {\hat J}~\lambda + {\rm det} {\hat J}=0
\end{equation}
Near the hyperbolic fixed point ($\theta_{\infty},s_{\infty}$)$=$($0,0$), 
one obtains two real eigenvalues ($\lambda_1<1$, $\lambda_2>1$)
\begin{equation}
\lambda_{1,2}=1+
{{{\cal S}(\theta_{\infty})}\over 2}
%{{2K_B+g\mu_B H_{\Vert}}\over {2J}} 
\mp \sqrt{
{{{\cal S}(\theta_{\infty})}\over 2}
%{{2K_B+g\mu_B H_{\Vert}}\over {2J}
\left\lbrack
2+
{{{\cal S}(\theta_{\infty})}\over 2}
%{{2K_B+g\mu_B H_{\Vert}}\over {2J}}
\right\rbrack
}
\end{equation}
and the slopes, in the ($\theta,s$) phase space, of the 
orbit inflowing to ("1") and outflowing from ("2") the 
hyperbolic fixed point are, respectively
\begin{equation}
m_{1,2}=
\label{m1}
{{ 
{{\partial s_{n+1}}\over {\partial \theta_n}} 
}\over
{
\lambda_{1,2}- {{\partial s_{n+1}}\over {\partial s_n}} }
}{\Bigg\vert}_{~(0,0)}
%=
%{  {{2K_B+g\mu_B H_{\Vert}}\over J}\over {
%{{2K_B+g\mu_B H_{\Vert}}\over {2J}} \mp \sqrt{
%{{2K_B+g\mu_B H_{\Vert}}\over {2J}}
%\big(2+{{2K_B+g\mu_B H_{\Vert}}\over {2J}}\big)
%}
%}}
\end{equation}

Within this theoretical framework, the condition for surface
magnetic canting is that $m_s$, the (negative) slope 
of the boundary condition curve Eq.~(\ref{surface}), calculated 
in the hyperbolic fixed point 
\begin{equation}
m_s={{ds_1}\over {d\theta_1}}{\Big\vert}_{~(0,0)}
={{2(K_S-K_B)}\over J}
\end{equation}
should be smaller than $m_1$, the (negative) slope of the 
trajectory inflowing to the hyperbolic fixed point, see
Eq.~(\ref{m1}). Thus, the condition for surface magnetic 
canting turns out to be 
\begin{equation}
\label{slopes}
{{2(K_S-K_B)}\over J}<
  { 
{{\cal S}(\theta_{\infty})}\over {
{{{\cal S}(\theta_{\infty})}\over 2}
%{{2K_B+g\mu_B H_{\Vert}}\over {2J}} 
- \sqrt{
{{{\cal S}(\theta_{\infty})}\over 2}
\left\lbrack
2+
{{{\cal S}(\theta_{\infty})}\over 2}
%{{2K_B+g\mu_B H_{\Vert}}\over {2J}}
\right\rbrack
%{{2K_B+g\mu_B H_{\Vert}}\over 
%{2J}} (2+{{2K_B+g\mu_B H_{\Vert}}\over {2J}})
}}}
\end{equation}
Taking into account that $K_S<0$ and $K_B>0$, this equation 
can be rewritten as 
\begin{eqnarray}
{{\vert K_S \vert}\over {J}} &>& {{2K_B+g\mu_B H_{\Vert}}\over {4J}}
\sqrt{1+{{4J}\over {2K_B+g\mu_B H_{\Vert}}}}
\cr &-&{{2K_B-g\mu_B H_{\Vert}}\over {4J}}
\label{Noi}
\end{eqnarray}
which is readily seen to be completely equivalent to Eq.~(\ref{PPC}).
Thus, in the case of external magnetic field parallel to the film plane, 
for sufficiently high surface anisotropy ({\it i.e.},
$2|K_S|>J+g\mu_B H_{\Vert}$), the surface is always canted
whatever the value of $K_B$.
Otherwise, there is SMC for $|K_S|$ exceeding a threshold value 
which depends on $K_B$, $J$ and $H_{\Vert}$: see Eq.~(\ref{Noi}).

\begin{figure}
\centerline{
}
\caption{
Map phase portraits calculated from 
Eqs.~(\ref{mappabulk})(\ref{mappageneral}) 
using Hamiltonian parameters $J=1$, $K_B=0.1$, $K_S=-0.3$ 
for different values of a magnetic field $H_{\Vert}$ 
applied parallel to the film plane: 
(a) $g\mu_B H_{\Vert}=0$; (b) 0.05; (c) $0.15\overline{5}$ (threshold
value, see Eq.~(\ref{threshold}); (d) 0.3.
The dashed curve represents the boundary condition 
at the surface, Eq.~(\ref{surface}). 
Arrows denote inflowing and outflowing trajectories 
associated with hyperbolic fixed points.
}
\end{figure}

In Fig.~1 we report the different map phase portraits
obtained from Eqs.~(\ref{mappabulk})(\ref{mappageneral})
%using the parameters $K_S=-0.3$, $K_B=0.1$ and $J=1$ 
for different, increasing values of $H_{\Vert}$. 
The boundary condition curve -
which does not depend on the value of $H_{\Vert}$,
see Eq.~(\ref{surface}) -
is also reported (dashed line). 
In zero field, $H_{\Vert}=0$ (see Fig.~1a),
for the chosen values of the parameters, 
the condition, Eq.~(\ref{slopes}), for the slopes $m_s$ and $m_1$
calculated in the hyperbolic fixed point,
is satisfied: thus, the configuration with 
surface magnetic canting is the ground state.
As $H_{\Vert}$ is increased, the slope $m_1$ of the 
trajectory inflowing to ($\theta_{\infty},s_{\infty}$) $=$(0,0)
becomes more negative and $\theta_1$ decreases 
(see Fig.~1b).
Finally, for $H_{\Vert} \ge H_{C\Vert}$, where 
$H_{C\Vert}$ is a threshold value given by
\begin{equation}
\label{threshold}
g\mu_B H_{C\Vert}=  
{{\left\lbrack 2(K_B-K_S) \right\rbrack^2}\over
{J+ 2(K_B-K_S) }}-2K_B~,
\end{equation}
the boundary condition curve
is no more intersected: thus, the uniform configuration 
with all spins parallel to the film plane becomes 
energetically favoured (see Figs.~1c,d).

Let us now consider a semi-infinite ferromagnetic film ($J>0$)
with competing surface and bulk anisotropies ($K_S<0$, $K_B>0$) and 
with an external magnetic field applied perpendicularly to the
film plane.\cite{nota2} The energy is
\begin{eqnarray}
\label{modelperp}
E&=&-J\sum_{n=1}^{\infty} \cos(\theta_n-\theta_{n+1})
+ K_S~\sin^2 \theta_1 
\cr &+&K_B \sum_{n=2}^{\infty} \sin^2 \theta_n
-g\mu_B H_{\perp} \sum_{n=1}^{\infty} \sin\theta_n
\end{eqnarray}
The map equations are now ($n \ge 1$)
\begin{eqnarray}
\label{mappabulkperp}
J s_{n+1}&=&J s_n+K_B \sin(2\theta_n)
-g\mu_B H_{\perp} \cos\theta_n
\cr
\theta_{n+1}&=&\theta_n+\sin^{-1}(s_{n+1})
\end{eqnarray}
while the boundary condition at the surface plane
turns out to be the same as in the case of in-plane
magnetic field, Eq.~(\ref{surface}).
The Jacobian determinant takes the same form as 
in Eq.~(\ref{Jacobian}):
the only difference is that 
now ${\cal S}(\theta_n)=
{{2K_B}\over J} \cos(2\theta_n)
+{{g\mu_B H_{\perp}}\over J}\sin\theta_n$.
In contrast with the case of in-plane field, now 
the hyperbolic fixed point 
($\theta_{\infty},s_{\infty}$) is
characterized by an angle, $\theta_{\infty}$, 
which depends on the field intensity 
\begin{equation}
\label{tinfinity}
\theta_{\infty} = \left\{
\begin{array}{l}
	\sin^{-1}
\left\lbrack
{{g \mu_B H_{\perp}}\over {2K_B}}
\right\rbrack
\; \; \mbox{,~for} \; 
g \mu_B H_{\perp}< 2K_B 
\; ,
	\cr
{{\pi}\over 2}
\; 
\; 
\mbox{~~~~~~~~~~~~~~~~~~,~for}\; 
g \mu_B H_{\perp} \ge 2K_B 
\; 
\end{array}\right.
\end{equation}
and $s_{\infty}=0$.
In Fig.~2 we report different map phase portraits, 
obtained from Eq.~(\ref{mappabulkperp}) 
%using the parameters $K_S=-0.3$, $K_B=0.1$ and $J=1$ 
calculated for different, increasing values of
$H_{\perp}$. The boundary condition curve Eq.~(\ref{surface}),
independent of the value of $H_{\perp}$,
is also reported (dashed line). 
It is apparent from the map topology 
(see Figs.~2a,b) that for $0< H_{\perp}<H_{C\perp}$, 
where $g \mu_B H_{C \perp}=2K_B$
is a threshold field value,
one does have surface magnetic canting
even in the case that SMC is absent  
for zero field.
In contrast, for $H_{\perp} \ge H_{C \perp}$ 
(see Figs.~2c,d), no intersection 
is possible between the boundary condition curve 
and the trajectory inflowing in the hyperbolic fixed point
($\theta_{\infty},s_{\infty}$)=(${{\pi}\over 2},0$):
thus, the uniform configuration with all spins 
perpendicular to the film plane becomes the ground state.

\begin{figure}
\centerline{
}
\caption{
Map phase portraits calculated from Eq.~(\ref{mappabulkperp}) 
using Hamiltonian parameters $J=1$, $K_B=0.1$, $K_S=-0.3$ 
for different values of a magnetic field $H_{\perp}$ 
applied perpendicularly to the film plane: 
(a) $g\mu_B H_{\perp}=0$; (b) 0.10; (c) 0.20 (threshold
value, $g\mu_B H_{C\perp}=2K_B$); (d) 0.3.
The dashed curve represents the boundary condition 
at the surface, Eq.~(\ref{surface}).
Arrows denote inflowing and outflowing trajectories 
associated with hyperbolic fixed points.
}
\end{figure}

\begin{figure}
\centerline{
}
\caption{
(a) Equilibrium configurations of the semi-infinite film
with Hamiltonian parameters $J=1$, $K_B=0.1$, $K_S=-0.3$,
as calculated from the map equations 
Eqs.~(\ref{mappabulk})(\ref{mappageneral})
and boundary condition at the surface Eq.~(\ref{surface}).
The different curves refer to selected values of
$g\mu_B H_{\Vert}=$ 0, 0.10, 0.15, and $H_{\Vert}>H_{C\Vert}$,
see Eq.~(\ref{threshold}), with the field increasing from top
to bottom. The lines are guides to the eye.
(b) The same as in (a), but for magnetic 
field applied perpendicularly to the film plane.
The different curves refer to selected values of
$g\mu_B H_{\perp}=$ 0, 0.10, 0.15, 0.19, 
and $H_{\perp}\ge H_{C\perp}$ (where $g\mu_B H_{C\perp}=2K_B=0.2$),
with the field increasing from bottom to top.
}
\end{figure}

In Fig.~3a,b we report the zero-temperature 
equilibrium configurations of the semi-infinite film 
obtained via the map method for different values of
the intensity of an external magnetic field,
applied in-plane or perpendicularly to the plane,
respectively. In both cases we have chosen Hamiltonian
parameters which, in zero field, give rise to surface 
magnetic canting. 
We observe that as the in-plane field $H_{\Vert}$ is increased
(see Fig.~3a from top to bottom), the surface angle 
$\theta_1$ decreases until, for $H_{\Vert}\ge H_{C\Vert}$, 
a uniform configuration with all spin in-plane is obtained.
In contrast, in the case of increasing perpendicular field (see Fig.~3b
from bottom to top), the surface angle $\theta_1$ increases 
and so does the bulk angle $\theta_{\infty}$, defined 
in Eq.~(\ref{tinfinity}). As $g\mu_B H_{\perp}$ reaches the 
threshold value $2 K_B$, a uniform configurations 
with all spins perpendicular to the film plane is obtained.

It is worth noting that by the map method it is possible
to calculate, with any desired accuracy, 
the equilibrium configuration even
in the case of a film with a finite number $N$ of spins,
provided that the boundary condition on the 
second surface,\cite{JMMM,Review}
%\begin{equation}
$s_N=-{{K_S-K_B}\over J} \sin(2\theta_N)$, 
%\end{equation}
is taken into account, in addition to Eq.~(\ref{surface}).\cite{nota4}

Finally, we observe that the approach described here 
can be used to investigate the zero-temperature 
magnetic properties of the Fe/Gd film,\cite{exptPP} provided that 
the model (\ref{model}) is modified in order to
account for the antiferromagnetic coupling between the Fe 
surface plane and the Gd underlying one.\cite{Ucraini}
However, owing to the absence of frustration effects,
only qualitative modifications to the results
obtained here are expected. 
Moreover, it is worth noting that 
the non-uniform ground state is not affected by 
a surface enhancement of the exchange constant (which is 
present in the Fe/Gd film, since the 1.5 monolayers of Fe 
order at a Curie temperature substantially higher than Gd).
In fact, since at $T=0$ the magnetization of each plane 
takes the saturation value, the surface exchange only keeps 
the spins on the surface parallel to each other, without modifying 
their orientation with respect to the spins on the underlying 
plane: {\it i.e.}, at $T=0$ the possibility of a non-uniform
ground state only arises from the competition between 
surface and bulk anisotropies.\cite{Review}
In contrast, the interpretation of the quite interesting finite 
temperature properties of Fe/Gd films, 
like the two-step reorientation transition 
experimentally observed by Arnold {\it et al.},\cite{exptPP} 
would require not only an improvement of model (\ref{model}) 
but also a much more refined analysis. In fact,
at $T \ne 0$ both the modulus and the orientation of the
magnetization vary with the plane index, so that the dimensionality 
of the map increases and the equilibrium configuration 
of the film cannot generally be calculated without 
resorting to some approximation.\cite{PLA} 

In conclusion, a mean-field theoretical method,
where the equilibrium properties of a magnetic film are formulated
in terms of an area-preserving map and the surfaces are introduced 
as appropriate boundary conditions, was exploited in this paper.
In particular, the model of a semi-infinite Heisenberg ferromagnet 
with competing bulk and surface anisotropies was investigated 
when an external magnetic field is applied parallel or 
perpendicular to the film surface. In both cases, the condition 
for surface magnetic canting as well as the 
zero-temperature equilibrium configuration  
were easily determined by such a theoretical approach. 

\begin{acknowledgments}
This paper is dedicated to our colleague and friend Matteo Amato.
\end{acknowledgments}


\begin{references}
\bibitem[3]{nota1}
See  Eq. (10) of Ref. \onlinecite{PLA} or Eq. (53) 
of Ref. \onlinecite{Review}. (Actually, this equation  
was firstly reported - with a misprint - in Ref. \onlinecite{JMMM}.)
In these papers, SMC was denoted as 
"surface magnetic reconstruction".
It should be noted that these equations differ slightly from 
Eq. (\ref{PPC}) or, equivalently, from Eq. (\ref{Noi}) 
of the present Brief Report just because, 
in Refs. \onlinecite{PLA} and \onlinecite{Review}, 
the bulk anisotropy was assumed to be present also on the
$n=1$ plane, in addition to the surface anisotropy.
\bibitem[10]{nota2}
Such a model was not considered by the authors in Ref. 1.
\bibitem[11]{nota3}
In practice, such a linearization procedure is  
equivalent to the second order $\theta_n$-expansion of the energy
in Eq.~(\ref{model}), performed by Popov and Pappas.\cite{PP}
\bibitem[12]{nota4}
In the finite film case, the non-uniform 
ground state configuration is found to be associated with elliptic 
trajectories.\cite{JMMM,Review}
\end{references}
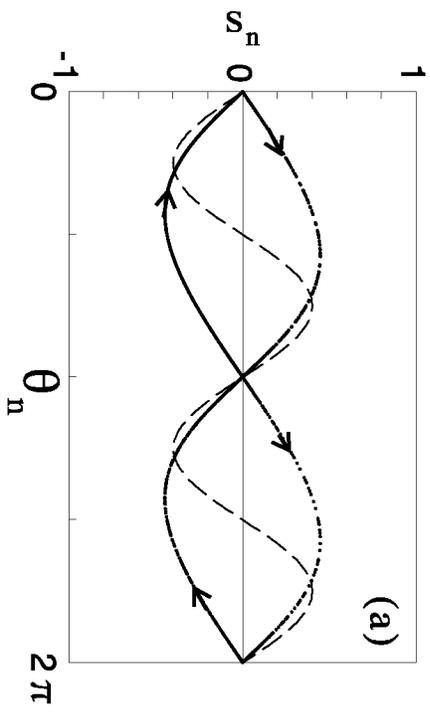
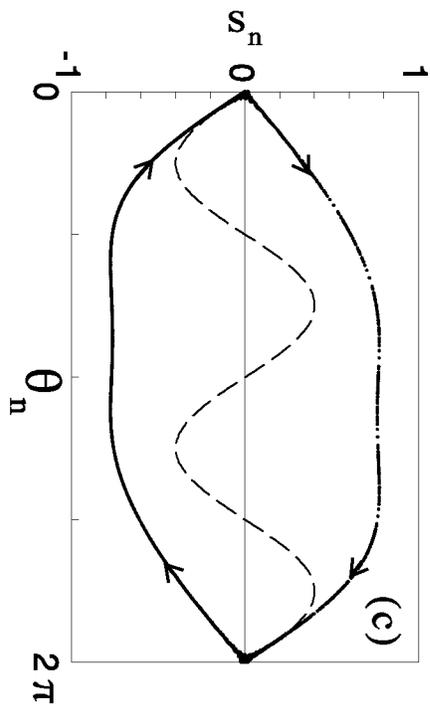
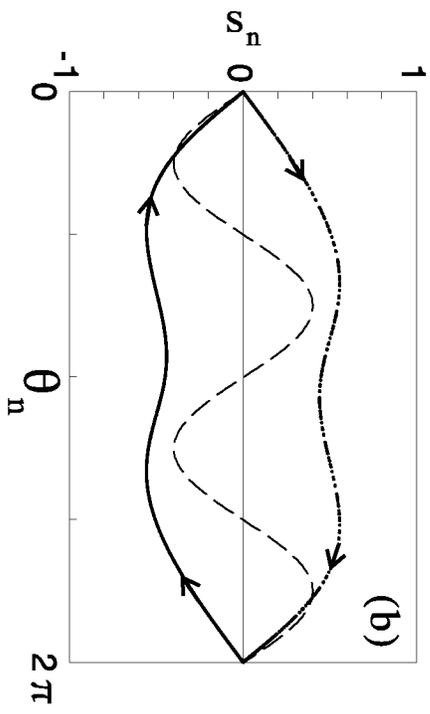
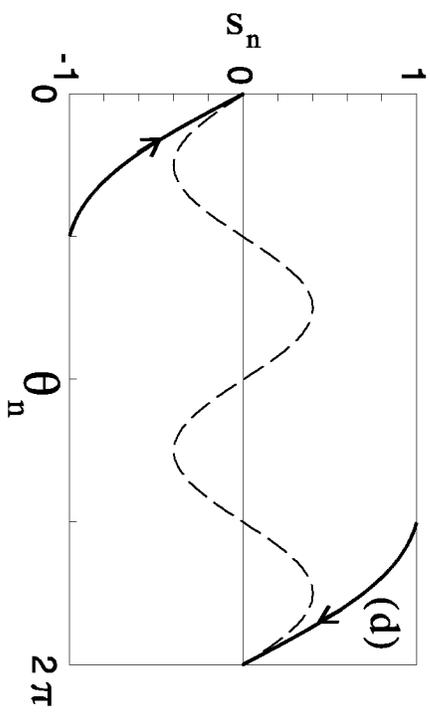
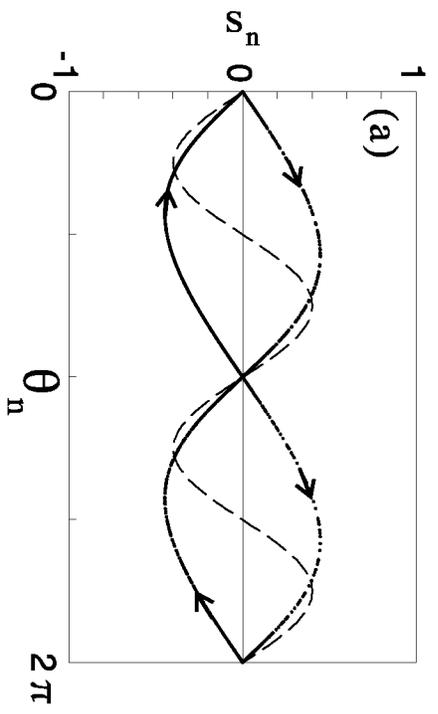
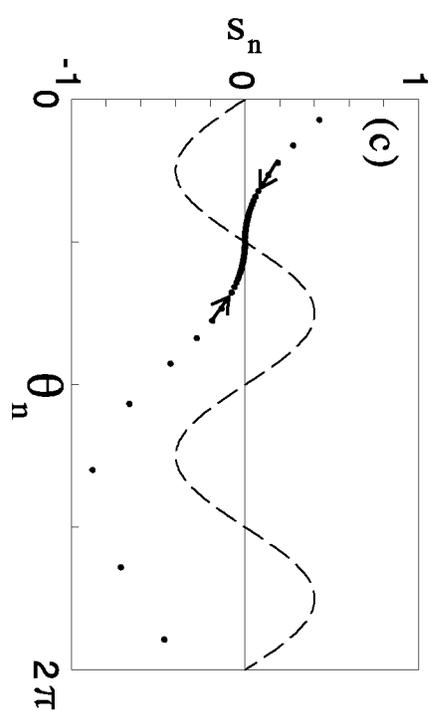
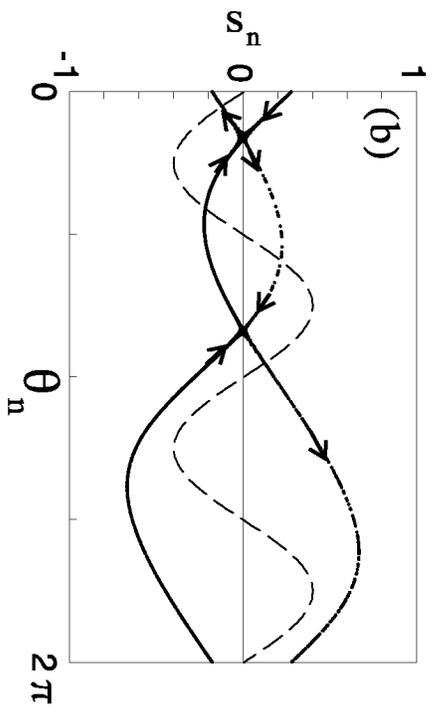
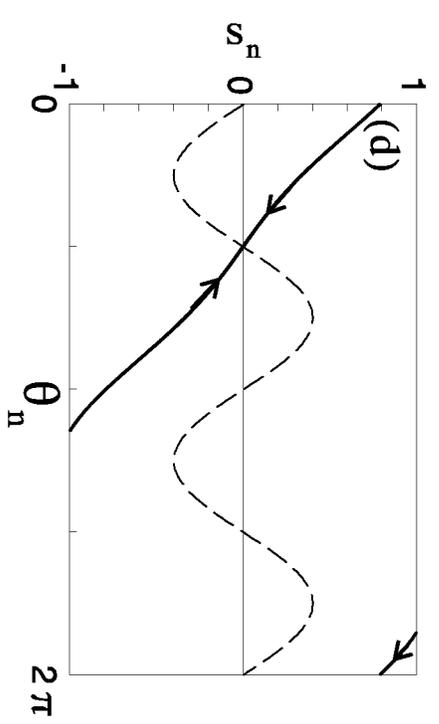
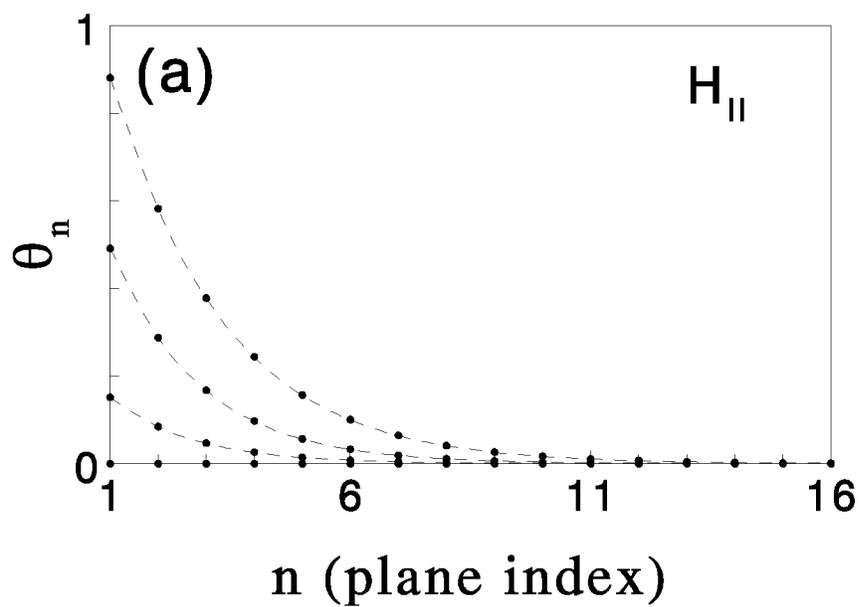
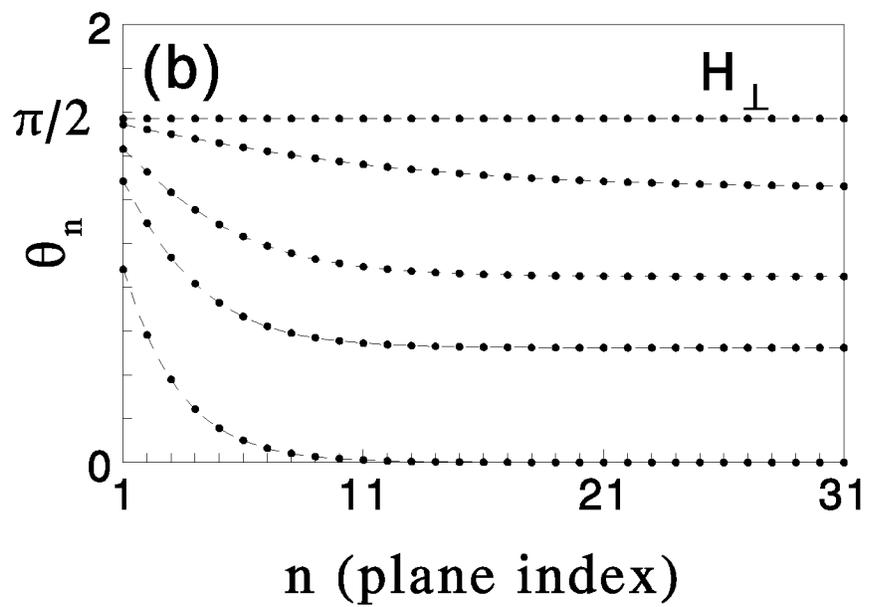
\end{document}